# A Deep Convolutional Neural Network Model for improving WRF Forecasts

Alqamah Sayeed[a], Yunsoo Choi[*,a], Jia Jung[a], Yannic Lops[a], Ebrahim Eslami[a, b], Ahmed Khan Salman[a]

[a]Department of Earth and Atmospheric Sciences, University of Houston, TX 77004
[b]Houston Advanced Research Center, The Woodlands, TX 77381
*corresponding author, ychoi23@central.uh.edu

*Abstract-*Advancements in numerical weather prediction models have accelerated, fostering a more comprehensive understanding of physical phenomena pertaining to the dynamics of weather and related computing resources. Despite these advancements, these models contain inherent biases due to parameterization and linearization of the differential equations that reduce forecasting accuracy. In this work, we investigate the use of a computationally efficient deep learning method, the Convolutional Neural Network (CNN), as a post-processing technique that improves mesoscale Weather and Research Forecasting (WRF) one day forecast (with a one-hour temporal resolution) outputs. Using the CNN architecture, we bias-correct several meteorological parameters calculated by the WRF model for all of 2018. We train the CNN model with a four-year history (2014-2017) to investigate the patterns in WRF biases and then reduce these biases in forecasts for surface wind speed and direction, precipitation, relative humidity, surface pressure, dewpoint temperature, and surface temperature. The WRF data, with a spatial resolution of 27 km, covers South Korea. We obtain ground observations from the Korean Meteorological Administration station network for 93 weather station locations. The results indicate a noticeable improvement in WRF forecasts in all station locations. The average of annual index of agreement for surface wind, precipitation, surface pressure, temperature, dewpoint temperature and relative humidity of all stations are 0.85 (WRF:0.67), 0.62 (WRF:0.56), 0.91 (WRF:0.69), 0.99 (WRF:0.98), 0.98 (WRF:0.98), and 0.92 (WRF:0.87), respectively. While this study focuses on South Korea, the proposed approach can be applied for any measured weather parameters at any location.

I. INTRODUCTION

The atmosphere sciences, particularly weather forecasting, have at their disposal a deluge of data from space, in-situ monitoring, and numerical simulations. These diverse data sources offer new opportunities, still largely underexploited, to improve our understanding, modeling, and reconstruction of geophysical dynamics. A number of academic studies, devoted to the problem of forecasting difficult-to-retrieve weather events and their associated uncertainties, typically employ weather forecasting techniques that fall into three main categories: numerical weather predictions (NWP), statistical forecasting, and artificial intelligence (AI - forecasting). Dynamical (physical) models such as the Weather Research and Forecasting (WRF) model use meteorological and topological information to determine the mesoscale weather parameters of a specific region [1], and statistical methods mainly use historical meteorological data to forecast the future state of the weather [2]–[6].

To obtain the various meteorological parameters, NWP models generally entail the parameterization of physical phenomena using initial and boundary conditions and a series of partial differential equations [7]. Unfortunately, despite advancements in these models, resolving horizontal resolutions through insufficient physical parameterization has led to unreliable weather forecasts [8]. NWPs are also computationally expensive, particularly with regard to fine-resolution forecasting [6]. In addition, because of the misrepresentation of unresolved small-scale features or neglected physical processes, parts of numerical models are represented by empirical sub-models or parameterizations [9]–[11], which tend to simplify involved physics that may lead to uncertainties in forecasting.

Unlike NWPs, statistical models require a large amount of historical data and completely neglect the physics of the atmosphere; thus, they do not consider meteorology [5], [12]. Since statistical methods are easily implemented and less computationally intensive than NWPs, they are popular among researchers. Nevertheless, owing to the scarcity of representing complex meteorological phenomena and non-linear patterns in the training data, statistical models are unreliable and inaccurate for forecasting extreme weather episodes, which exacerbate for long-range forecasting.

Because of the chaotic nature of the weather system, errors in weather forecasting are unavoidable but quite often significant regardless of the implemented modeling approach. The parametrization and linearization of differential equations lead to biases, which increases at every step of space and time in a numerical model. Overcoming these limitations still remains a challenging task. In the past several decades, the volume and quality of observations have increased dramatically, particularly thanks to remote sensing. At the same time, new developments in machine learning (ML), particularly deep learning (DL) [13], have demonstrated impressive capabilities at reproducing complex spatiotemporal processes [14] by efficiently using an enormous amount of data, thus creating a path for their use in the atmospheric sciences.

Researchers have applied various ML algorithms in a variety of fields in the earth and atmospheric sciences, including air quality forecasting [2]–[4], [6] and hurricane tracking [15]. ML has also been applied to nowcasting based on real observations such as the sea surface temperature [16] and precipitation [17]. In this study, we apply an alternative approach: a fully data-driven framework that combines a deep neural network and physical models that simulate the dynamics of a complex weather system. We have developed a weather-AI as a gridded real-time weather forecasting model that reduces the model-measurement error of the WRF model. The system, using a convolutional neural network algorithm [18], post-processes and bias-corrects the WRF output (observation network of the 24-hour forecast) in real-time at each grid linked to a station location.

## II. METHODOLOGY

The algorithm is divided into two sections: i) hourly forecast by a WRF model and ii) a deep CNN model that reduces uncertainty and improves forecasting accuracy. Fig. 1 shows the process flow diagram for the Weather-AI model.

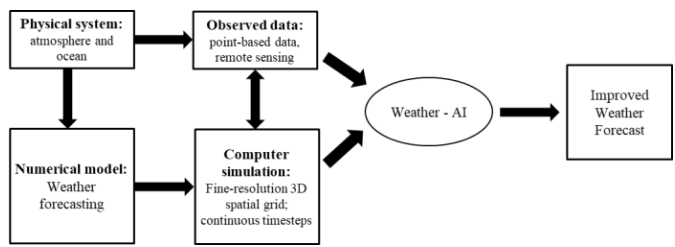

**Figure 1.** Process flow for the Weather - AI model in bias correcting WRF forecasts. A Weather-AI model uses historical simulation by a numerical model (WRF) and uses the actual observation to understand the biases. The process is called training an AI model. Once a model is trained, it is used to forecast unseen scenarios.

### A. WRF Configuration:

WRF v3.8 covers the eastern part of China, the Korean Peninsula, and Japan, with a 27 km horizontal grid spacing for the years 2014 to 2018. Detailed configurations of the WRF model are available in [19].

### B. Deep Convolutional Neural Network

The deep architecture of the convolution neural network used in this study is similar to the model in [6]. The model entails five one-dimensional convolutional layers, a fully connected layer, and an output layer. Each convolutional layer, with 32 filters, is activated by the rectified linear unit. The input for the first layer consists of various hourly meteorological parameters extracted from the WRF model (Table T1 in the Supplementary Document lists all the WRF meteorological parameters used as input). The convolutions are applied to the input features with the elements of a randomly initialized kernel (with a kernel window size of 2 x 1). The feature maps are obtained through the output of the first layer, then used as input for the second layer. The same process is applied in the succeeding layers. The output of the fifth convolutional layer is then passed to the fully connected layer, which contains 264 nodes (neurons). The hourly output is obtained at the last layer (output layer). A deep CNN, like any neural network, is an optimization problem that attempts to minimize the loss function. This study uses a loss function based on the index of agreement (IOA) [20], developed by [21].

### C. Data Preparation and Model Training

We obtained observed meteorology from the 93 Automated Synoptic Observing System (ASOS) stations operated by the Korea Meteorological Administration (KMA) for the years 2014 to 2018 across South Korea. Fig. 2 displays the location of all the meteorology monitoring stations in the country. The meteorological parameters obtained from these stations were wind speed, wind direction, precipitation, relative humidity, temperature, dewpoint temperature, and surface pressure.

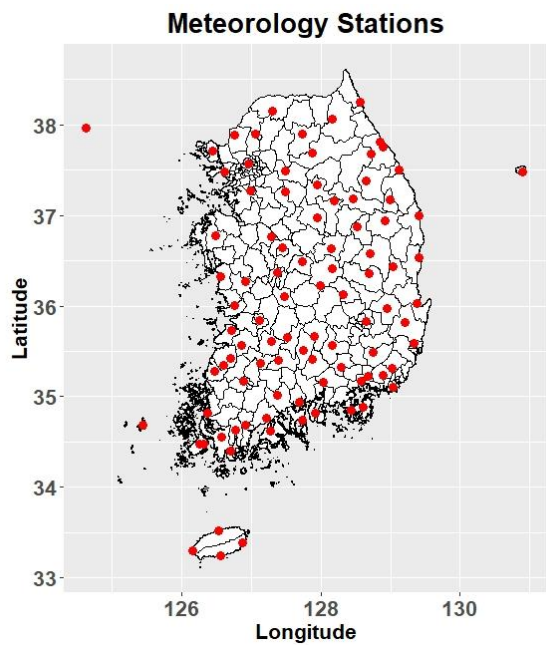

**Figure 2.** Map of South Korea with the location of the meteorology stations used in the Weather-AI model for bias-correcting WRF weather forecasts.

Upon completion of the WRF run, we identified the closest WRF grid to each station, to which we assigned a grid point (Table T2 in the Supplementary Document), and then extracted hourly meteorology at each grid point (Table T1 in the Supplementary document). After acquiring hourly meteorological fields from the output of the WRF model, we prepared the input for each station in the form of a two-dimensional matrix in which each column represented a specific meteorology parameter and each row represented hourly values. As each column represented a specific meteorological parameter, it displayed a range of values. To establish uniformity over all inputs, we normalized each column between 0 and 1 with a global minimum and maximum

[6]. The output dataset consisted of the hourly observed meteorology. To construct a matrix for training/testing a generalized deep CNN model across the spatial domain, we then combined all stations data row-wise and further split the training dataset into a 50-50 ratio (randomly) for training and validation. Then, we trained the model for three years (i.e., 2014 to 2017) and evaluated it for the year 2018 (Note: We did not use 2018 in the model training). We trained a separate model for each of the observed meteorological parameters.

*1) Special Case: Precipitation model*

Forecasting the amount of hourly rainfall for a specific region requires complex physics and chemistry pertaining to atmospheric conditions. Thus, we divided the forecasting of rainfall into two sections: a classification model (Rain-CM: Rain Classification model) that identified rain hours; and an hourly quantity prediction model (Rain-RM: Rain Regression Model). The two models are combined to forecast the hourly and daily accumulated total rainfall (in mm).

The Rain-CM model is similar to the model discussed in previous sections but differs in its output, which consists of 0's for no rain and 1's for rain hours. The data setup of the Rain-RM model differed slightly from that of the models discussed in this study. The output consisted of observed 24 hourly rain amounts (in mm) arranged in rows, and the inputs consisted of the first hour of a day (0000 UTC) forecasted meteorology and forecasted 24-hourly rain amounts (in mm) by the WRF. Therefore, each row in the setup consisted of daily values instead of hourly values.

### III. RESULTS AND DISCUSSION

For the Weather-AI model, we obtained the following meteorological parameters: wind speed, wind direction, temperature, pressure, dewpoint temperature, relative humidity, vapor pressure, and precipitation at the surface. We then applied this model, and the WRF model, to obtain forecasts for all of 2018.

*A. Wind Speed and Direction:*

Fig. 3 shows the performance of the WRF model (Fig. 3a) and the Weather-AI model (Fig. 3b) for each station in terms of IOA. The Weather-AI models show an average increase of 27% in IOA for all stations; IOA increased from 0.67 (correlation = 0.66) for the WRF model to 0.85 (correlation = 0.75) for the Weather-AI model. Overall, the Weather-AI model improved forecasting for all stations, with more than two-thirds (64 out of 93) of the stations showing an IOA increase greater than 20% (Fig. S1 shows the percentage change in the IOA at all stations).

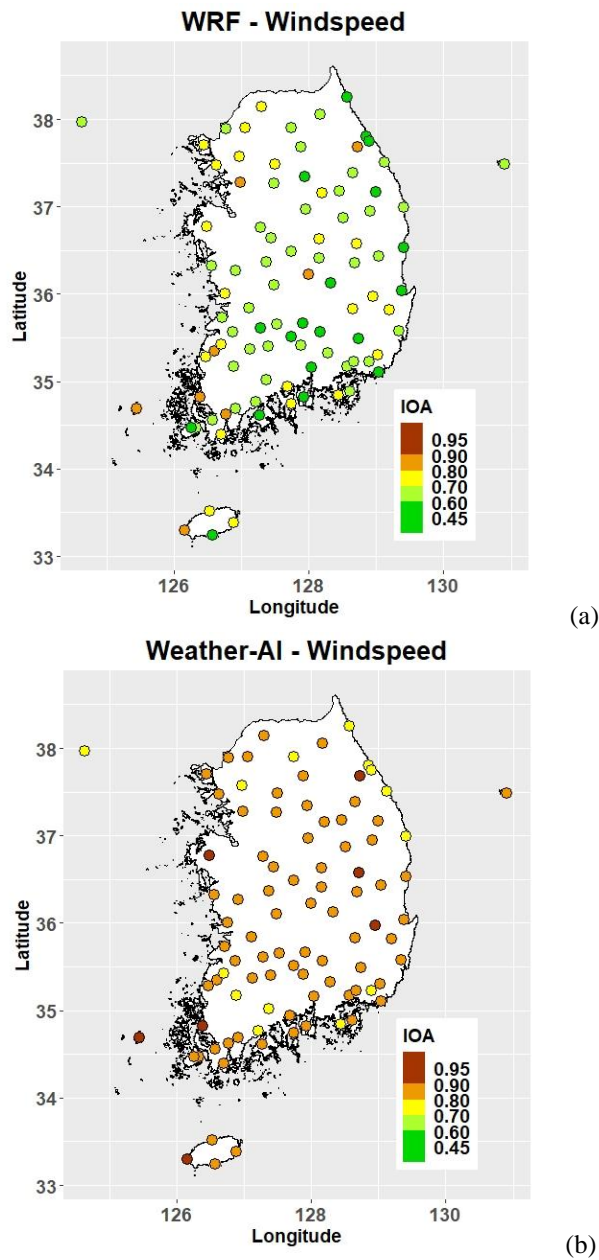

**Figure 3.** Station-wise IOA comparison of wind speed for a) WRF and b) Weather-AI models.

Fig. 4 shows Taylor diagrams (separated by month) comparing the performance of the two models for all stations combined. The figure shows that the model closest to the observed point on the diagram performs the best [21], demonstrating the superior performance of the Weather-AI model in all months. Although the root mean squared error (RMSE) for the WRF varied each month and was larger in the cold months, the RMSE for the Weather-AI remained constant at 1 m/s. Similarly, while the standard deviation (SD) and correlation of the WRF varied each month, those of the Weather-AI remained stable throughout the year. From Fig. 4, one can conclude that seasonality does not affect the forecasting of Weather-AI for wind speed.

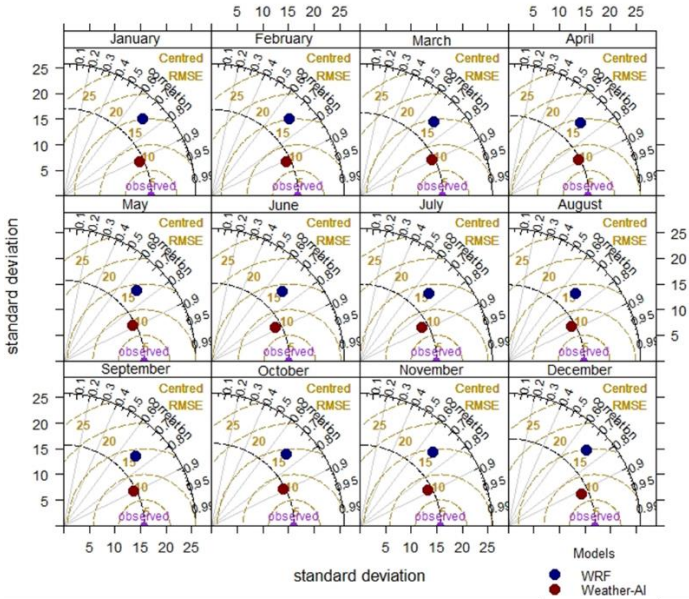

**Figure 4.** Taylor diagram of each month comparing the performance of the WRF forecast and Weather-AI bias-correction for wind speed.

Predicting the wind direction is challenging because of its circular nature. To do so, we first predict u and v components of winds and calculate the direction. To evaluate the performance of the wind direction, all the predictions that are in the bin of ±45° from observed values are treated as true predictions, and all other values are treated as false predictions. Hence, categorical statistic evaluations, in this case, are as follows:

$$HR_{wd}, \text{Hit Rate} = \frac{\text{No. of hours when both predictions are in the range of } \pm 45^0 \text{ from observed values}}{\text{Total no. of hours of observations}}$$

$$FAR_{wd}, \text{False Alarm Rate} = \frac{\text{No. of hours when predictions are not in the range of } \pm 45^0 \text{ from observed values}}{\text{Total no. of hours of observations}}$$

The $HR_{wd}$ for all stations combined for the Weather-AI was 54.83% and $HR_{wd}$ for the WRF was 52.16%.

Fig. 5a shows the yearly time series of wind speed and Fig. 5b shows the wind direction at station 115. This station is unique because it is situated near the southeastern coast of a small island, Ulleng-do (120 km east of the Korean Peninsula). The WRF model significantly overpredicted wind speeds during the cold months (Fig. 5a). As summer approached, its performance improved (also shown in Fig. 4), with the most dramatic improvement in the JJA season. The Weather-AI model was able to reduce the seasonal biases of the WRF, out-performing it in all months for predicting wind speed and more accurately predicting the wind direction (Fig. 5b). Furthermore, the model significantly improved the wind direction predictions by successfully predicting dominant southwestern and northeastern wind directions.

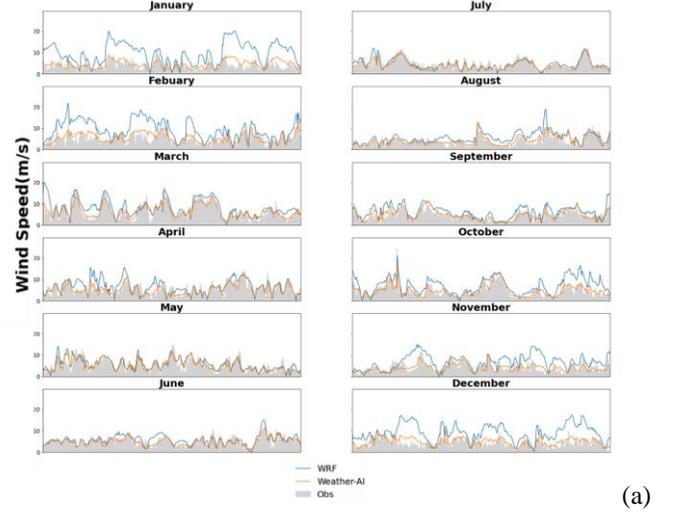

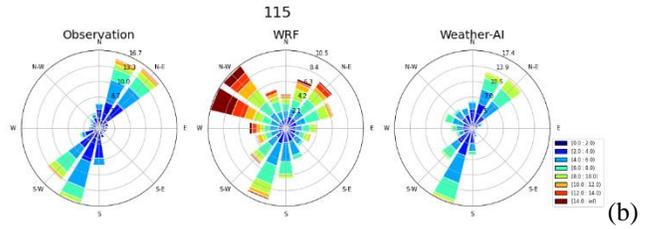

**Figure 5.** a) Hourly wind-speed time-series for station 115 for the year 2018. Each subplot represents a month of the year; the X-axis represents hours of the day and the Y-axis the wind speed in m/s. b) Polar plot of hourly wind direction in 2018.

*B. Precipitation*

Precipitation forecasting consisted of two models. Therefore, we used different techniques to evaluate them. We evaluated Rain-CM based on categorical statistics, that is, the hit rate (HR) and the false alarm rate (FAR), defined as follows:

$$HR_{rain}, \text{HR Rain Condition} = \frac{\text{No. of hours when both prediction and observation are a rain hour}}{\text{Total no. of hours when observation is a rain hour}}$$

$$FAR_{rain}, \text{FAR Rain Condition} = \frac{\text{No. of hours when prediction is a no rain and observation is a rain hour}}{\text{Total no. of hours when observation is a rain hour}}$$

$$HR_{no\text{-}rain}, \text{HR No-Rain Condition} = \frac{\text{No. of hours when both prediction and observation are no rain hour}}{\text{Total no. of hours when observation is a no rain hour}}$$

$$FAR_{no\text{-}rain}, \text{FAR No-Rain Cond.} = \frac{\text{No. of hours when prediction is rain observation is no rain hour}}{\text{Total no. of hours where observation is a no rain hour}}$$

**Table 1a.** Categorical evaluation of the rain classification for the WRF model.

|  | Observed Rain Hours | | Observed No Rain Hours | |
|---|---|---|---|---|
| **Predicted Rain** | 44058 | $HR_{rain} = 0.84$ | 122670 | $FAR_{no-rain} = 0.16$ |
| **Predicted No Rain** | 8256 | $FAR_{rain} = 0.16$ | 638654 | $HR_{no-rain} = 0.84$ |
| **Total Hours** | 52314 | | 761324 | |

**Table 1b.** Categorical evaluation of the rain classification for the Weather-AI (Rain-CM) model.

|  | Observed Rain Hours | | Observed No Rain Hours | |
|---|---|---|---|---|
| **Predicted Rain** | 47214 | $HR_{rain} = 0.90$ | 117572 | $FAR_{no-rain} = 0.15$ |
| **Predicted No Rain** | 5100 | $FAR_{rain} = 0.10$ | 643752 | $HR_{no-rain} = 0.85$ |
| **Total Hours** | 52314 | | 761324 | |

Tables 1a and 1b show the HR and FAR of the WRF and Weather-AI models, respectively, for the year 2019 for all stations combined (observations with "NaN" values were removed). The Weather-AI Rain-CM model showed 7% and 1% improvement over the WRF model in the HR for rain and no-rain hours, respectively, and 37.5% and 6.25% decrease in the FAR for rain and no-rain, respectively.

After obtaining the predictions from the classification model, we applied the regression model (Rain-RM) to predict the hourly amount of precipitation. To merge both models and forecast rain more accurately, we converted all the non-rain hours from the Rain-CM to zero. The average IOA for all stations for hourly rain was 0.62 (WRF = 0.56) and the correlation was 0.51 (WRF = 0.43). According to Fig. 6, which presents a station-wise IOA comparison for hourly rain, 90% of the stations show an improved IOA, and 95% show an improved correlation for hourly rain.

The next step in rain forecasting was daily accumulated rainfall, calculated from the hourly rain predicted by the Rain-RM model. Fig. 7 represents a station-wise IOA comparison of the WRF and Weather-AI models. The average IOA and correlation of the Weather-AI model were 0.87 (WRF-0.86) and 0.79 (WRF-0.77), respectively. Fig. 8 shows a scatter plot for daily accumulated total rainfall forecasted by the WRF and Weather-AI models, and displays a significant reduction in under-predictions and over predictions of high values.

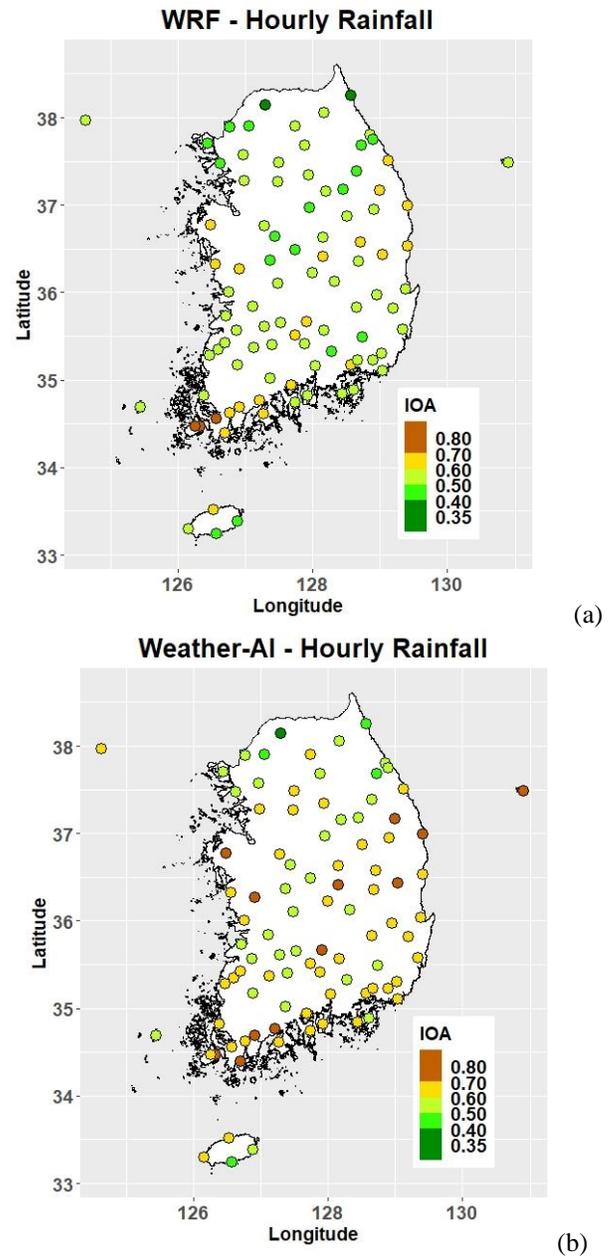

**Figure 6.** Station-wise IOA comparison of precipitation for a) the WRF and b) the Weather-AI models.

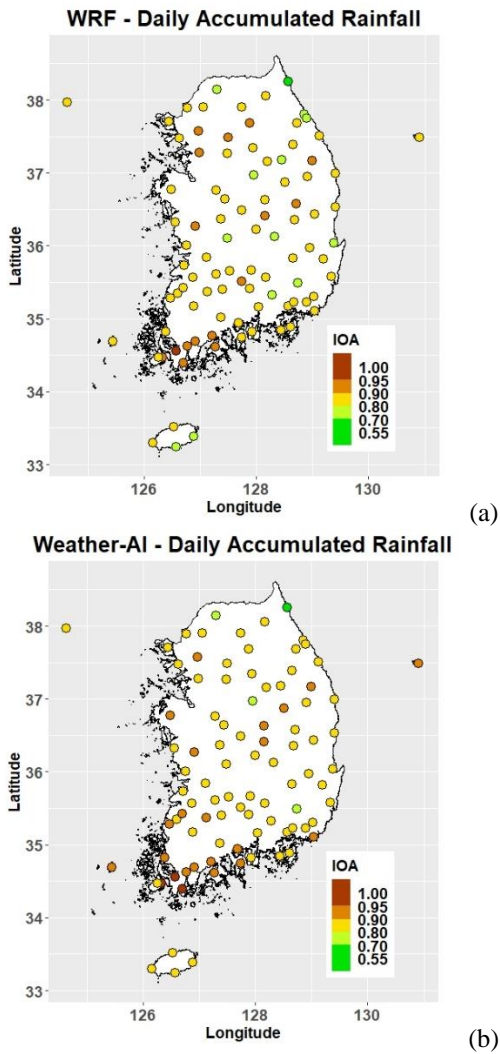

Figure 7. Station-wise IOA comparison of daily precipitation predictions by a) the WRF and b) Weather-AI models.

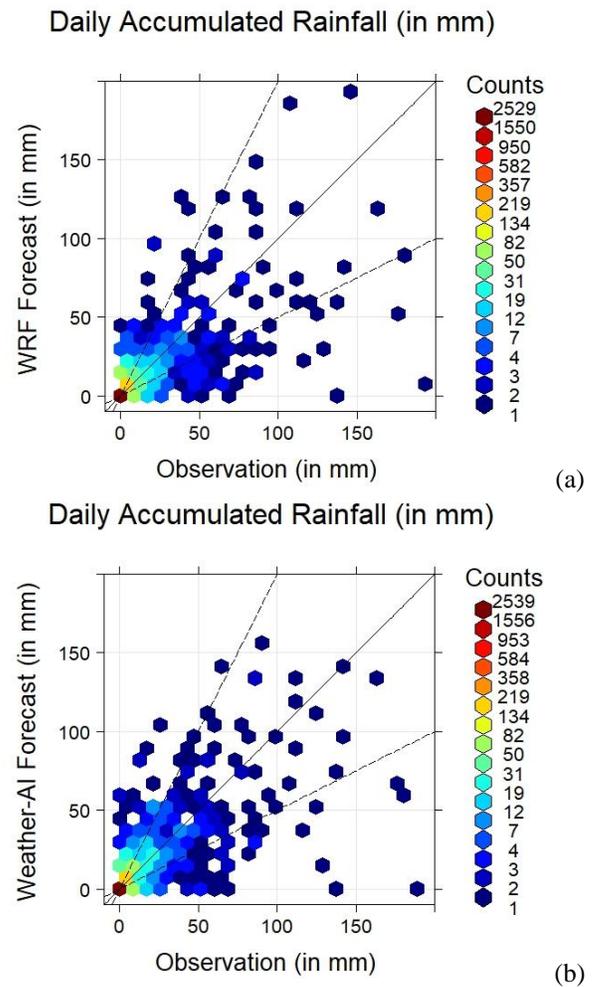

Figure 8. Scatter plot of daily accumulated rainfall prediction by a) the WRF and b) Weather-AI models. The X-axis represents observed rainfall in mm and the Y-axis represents predictions in mm.

*C. Other Meteorology:*

Fig. 9a and 9b present the station-wise IOA for forecasting hourly temperature 24 hours in advance for the WRF and Weather-AI models, respectively. Both models performed well in forecasting temperature, with an average IOA for all stations combined of 0.98 from the WRF and 0.99 from the Weather-AI models. The range of the IOA for the WRF was 0.92-0.99 and for the Weather-AI 0.98-0.99. Even though the temperature forecasts of the WRF were exceptionally accurate, those of the Weather-AI still showed improvements in all stations. A similar improvement occurred for the dewpoint temperature (Fig. 9c and 9d). A monthly Taylor diagram comparison of both models for temperature and dewpoint temperature forecasting are shown in Fig. 10a and 10b. Results have shown the RMSE and the SD from WRF were slightly larger during the DJF (December, January, and February) season with a weaker correlation. Whereas during the warmer months, WRF had smaller RMSE and SD with a higher correlation. In contrast, the Weather-AI generated more accurate predictions than the WRF for all months. The RMSE and SD did not vary or exceed 2°C for each month throughout the season.

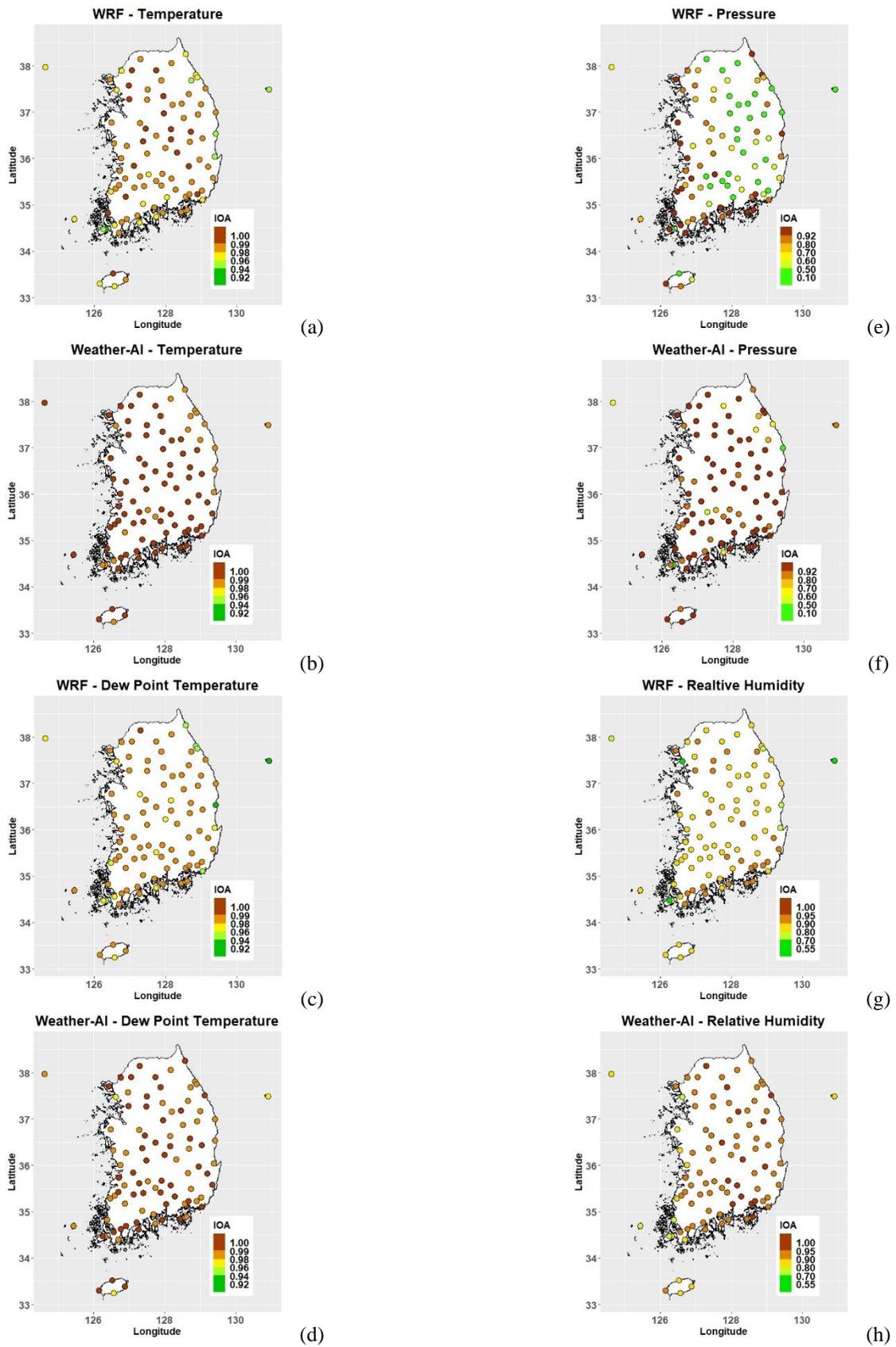

**Figure 9.** Station-wise IOA comparison of the forecasts of the WRF and Weather-AI models for temperature, dewpoint temperature, surface pressure and relative humidity.

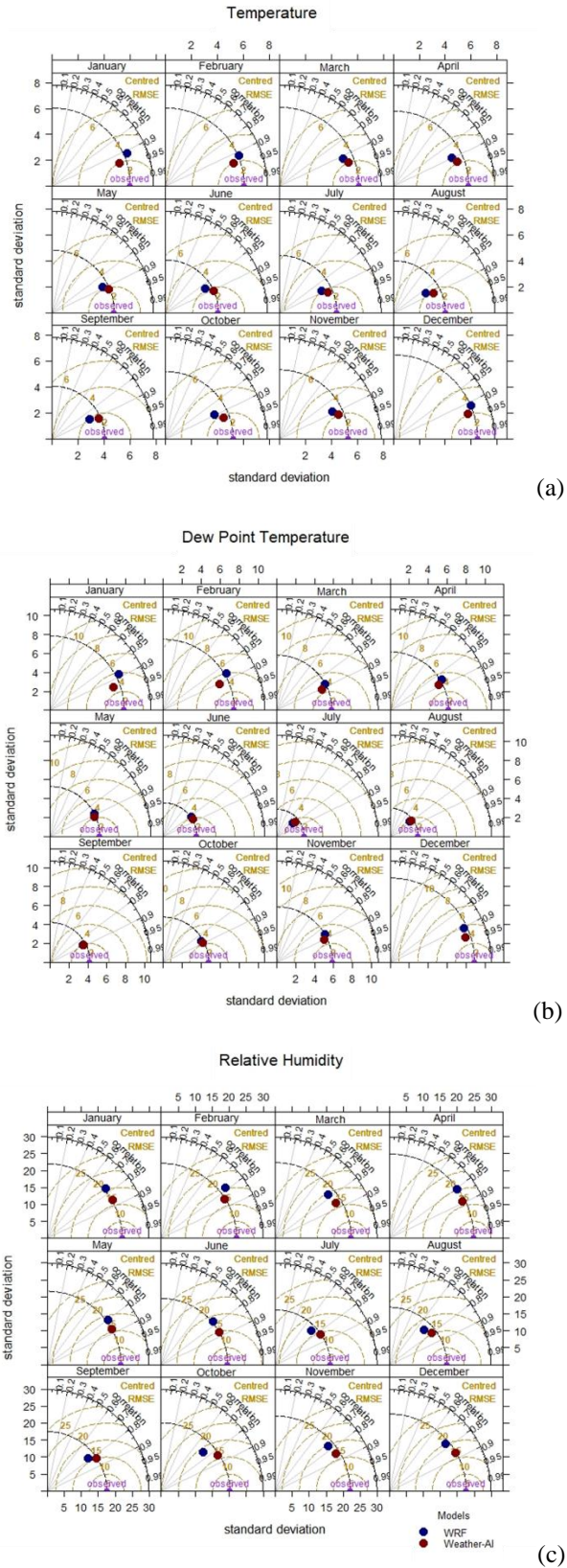

**Figure 10.** Taylor diagrams comparing the WRF and Weather-AI models for a) temperature, b) dewpoint temperature, c) surface pressure, and d) relative humidity for each month in 2018.

The IOA for the hourly surface pressure forecasts for 24 hours increased significantly, as shown in Fig. 9e and 9f. The average IOA of the WRF and Weather-AI models were 0.69 and 0.91, respectively. For several of the stations, the WRF produced uniform bias in forecasting surface pressure, which was adjusted by the Weather-AI (Fig. S2 in Supplementary Document). Since the bias from the WRF was uniform, the correlation was stronger for these stations, but the IOA was weaker. However, as the bias from the Weather-AI decreased, the IOA increased.

Fig. 9g and 9h show the yearly IOA of the hourly forecasts of relative humidity from the WRF (IOA-0.87) and Weather-AI (IOA-0.92) models, respectively. All, except for five (Station 169, 165, 129, 140, and 170), stations show improvement in the IOA. According to Fig. 10c, both the WRF and Weather-AI models display comparable performance for relative humidity; the forecasts of Weather–AI, however, are slightly more accurate than those of the WRF in all months.

## IV. CONCLUSION

In this paper, we developed and discussed a deep CNN model that reduced bias in an NWP model and significantly improved predictions. Although we retained the same model configuration, we developed several meteorology-specific models based on the target/output. The models showed improved predictions over the WRF model and significantly reduced bias.

The IOA for wind speeds from the Weather-AI model improved for all 93 stations in South Korea. Improvement fell within the range of 2.3 – 39.3%, with a mean of 17.83% in absolute terms. For wind direction, the predictions of the Weather-AI model improved in 52 out of 93 stations. Moreover, the performance remained consistent throughout the year. The Rain-CM improved the hit rate by 6% over the WRF model for the prediction of rain hours, but it remained the same as the WRF for the prediction of no-rain hours. The predictions of the hourly rainfall amount by the Rain-RM model improved in most of the stations; nevertheless, forecasting the absolute amount of hourly rainfall remains a challenge. Predictions of the daily accumulated rainfall amount showed a slight improvement in the IOA and a 2% improvement in the correlation. The forecasting of other meteorological parameters— temperature, dewpoint temperature, and relative humidity— also improved.

The Weather-AI model significantly improved the predictions of wind speed, relative humidity, and hourly precipitation. As WRF predictions were already relatively accurate, it did not show significant improvement in the predictions of temperature and dewpoint temperature. The forecasting of surface pressure from WRF contained a uniform bias in several stations that were corrected by the Weather-AI model. Even though the Weather-AI model was trained for South Korea, a similar model can be trained and reproduced for any location to forecast any number of meteorological parameters, and it is computationally fast. Although the AI model showed significant improvement over the WRF model, it does not cover WRF domains over the sea/ocean (because of the lack of observations). In addition, unlike the WRF and more advanced architectures of CNN, the Weather-AI model has no spatial gridded structure. Therefore, we need to develop AI models capable of spatial and temporal forecasting, specifically long-range forecasting, based on Weather-AI.

ACKNOWLEDGMENTS

This study was supported by the High Priority Area Research Seed Grant of the University of Houston. The computing system from UH HPE DSI was used for this study.

# Supplemental Materials

## Tables

**Table T1.** List of WRF meteorological parameters extracted from each grid point. (*WRF diagnostic variables)

| Symbol | Description | Units |
|---|---|---|
| P_HYD | Hydrostatic Pressure | Pa |
| Q2 | Water Vapor Mixing Ratio at 2m | kg/kg |
| T2 | Temperature at 2m | K |
| TH2 | Potential Temperature at 2m | K |
| PSFC | Surface Pressure | Pa |
| U10 | U Wind at 10 M | m/s |
| V10 | V Wind at 10m | m/s |
| QVAPOR | Water Vapor Mixing Ratio | kg/kg |
| QCLOUD | Cloud Water Mixing Ratio | kg/kg |
| QRAIN | Rain Water Mixing Ratio | kg/kg |
| SHDMAX | Annual Max Veg Fraction | |
| SHDMIN | Annual Min Veg Fraction | |
| SNOALB | Annual Max Snow Albedo in Fraction | |
| TSLB | Soil Temperature | K |
| SMOIS | Soil Moisture | m3/m3 |
| SH2O | Soil Liquid Water | m3/m3 |
| SFROFF | Surface Runoff | mm |
| UDROFF | Underground Runoff | mm |
| IVGTYP | Dominant Vegetation Category | |
| ISLTYP | Dominant Soil Category | |
| VEGFRA | Vegetation Fraction | |
| GRDFLX | Ground Heat Flux | W/m2 |
| ACGRDFLX | Accumulated Ground Heat Flux | J/m2 |
| ACSNOM | Accumulated Melted Snow | kg/m2 |
| SNOW | Snow Water Equivalent | kg/m2 |
| SNOWH | Physical Snow Depth | m |
| CANWAT | Canopy Water | kg/m2 |
| SSTSK | Skin Sea Surface Temperature | K |
| COSZEN | Cos of Solar Zenith Angle | |
| LAI | Leaf Area Index | m2/m2 |
| VEGF_PX | Vegetation Fraction for PX LSM | area/area |
| CANFRA | Satellite Canopy Fraction | |
| VAR | Orographic Variance | |
| F | Coriolis Sine Latitude Term | s-1 |
| E | Coriolis Cosine Latitude Term | s-1 |
| HGT | Terrain Height | m |
| RAINC | Accumulated Total Cumulus Precipitation | mm |
| RAINSH | Accumulated Shallow Cumulus Precipitation | mm |
| RAINNC | Accumulated Total Grid Scale Precipitation | mm |

| | | | |
|---|---|---|---|
| SNOWNC | Accumulated Total Grid Scale Snow and Ice | mm | |
| GRAUPELNC | Accumulated Total Grid Scale Graupel | mm | |
| HAILNC | Accumulated Total Grid Scale Hail | mm | |
| CLDFRA | Cloud Fraction | | |
| SWDOWN | Downward Short-Wave Flux at Ground Surface | W/m2 | |
| GLW | Downward Long-Wave Flux at Ground Surface | W/m2 | |
| SWNORM | Normal Short-Wave Flux at Ground Surface (Slope-Dependent) | W/m2 | |
| OLR | TOA Outgoing Long Wave | W/m2 | |
| ALBEDO | Albedo | | |
| ALBBCK | Background Albedo | | |
| EMISS | Surface Emissivity | | |
| NOAHRES | Residual of the NOAH Surface Energy Budget | W/m2 | |
| TMN | Soil Temperature at Lower Boundary | K | |
| XLAND | Land Mask | | |
| PBLH | PBL Height | m | |
| HFX | Upward Heat Flux at the Surface | W/m2 | |
| QFX | Upward Moisture Flux at the Surface | kg m-2 s- | |
| LH | Latent Heat Flux at the Surface | W/m2 | |
| SNOWC | Flag Indicating Snow Coverage | | |
| SR | Fraction of Frozen Precipitation | | |
| SST | Sea Surface Temperature | K | |
| Ue10* | U-wind in Earth Coordinate | m/s | |
| Ve10* | V-wind in Earth Coordinate | m/s | |
| WS* | Wind Speed | m/s | |
| WD* | Wind Direction | | |

**Table T2.** Table showing the Latitude, Longitude, and distance between station and WRF grid points.

| Station ID | Station Latitude | Station Longitude | WRF Latitude | WRF Longitude | Distance (in kms) |
|---|---|---|---|---|---|
| 100 | 37.68 | 128.72 | 37.59 | 128.67 | 10.40 |
| 101 | 37.90 | 127.74 | 37.86 | 127.74 | 4.68 |
| 102 | 37.97 | 124.63 | 37.87 | 124.58 | 12.07 |
| 104 | 37.80 | 128.86 | 37.83 | 129.00 | 12.91 |
| 105 | 37.75 | 128.89 | 37.83 | 129.00 | 12.94 |
| 106 | 37.51 | 129.12 | 37.58 | 128.99 | 14.72 |
| 108 | 37.57 | 126.97 | 37.62 | 127.10 | 13.07 |
| 112 | 37.48 | 126.62 | 37.38 | 126.47 | 17.69 |
| 114 | 37.34 | 127.95 | 37.36 | 128.04 | 8.18 |
| 115 | 37.48 | 130.90 | 37.51 | 130.87 | 4.00 |
| 119 | 37.27 | 126.99 | 37.37 | 127.10 | 14.75 |
| 121 | 37.18 | 128.46 | 37.10 | 128.34 | 13.72 |
| 127 | 36.97 | 127.95 | 36.86 | 128.02 | 13.76 |
| 129 | 36.78 | 126.49 | 36.88 | 126.47 | 11.58 |
| 130 | 36.99 | 129.41 | 37.07 | 129.28 | 15.32 |
| 131 | 36.64 | 127.44 | 36.62 | 127.39 | 4.67 |
| 133 | 36.37 | 127.37 | 36.37 | 127.39 | 1.43 |
| 135 | 36.22 | 127.99 | 36.11 | 128.00 | 11.80 |
| 136 | 36.57 | 128.71 | 36.60 | 128.63 | 7.27 |
| 137 | 36.41 | 128.16 | 36.36 | 128.00 | 14.60 |
| 138 | 36.03 | 129.38 | 36.07 | 129.53 | 14.32 |
| 140 | 36.01 | 126.76 | 35.88 | 126.77 | 13.59 |
| 143 | 35.83 | 128.65 | 35.85 | 128.60 | 5.32 |
| 146 | 35.84 | 127.12 | 35.88 | 127.07 | 6.14 |
| 152 | 35.58 | 129.33 | 35.59 | 129.20 | 12.13 |
| 155 | 35.17 | 128.57 | 35.11 | 128.57 | 6.80 |
| 156 | 35.17 | 126.89 | 35.14 | 126.76 | 12.81 |
| 159 | 35.10 | 129.03 | 35.09 | 129.18 | 13.20 |
| 162 | 34.85 | 128.44 | 34.86 | 128.56 | 11.67 |
| 165 | 34.82 | 126.38 | 34.89 | 126.45 | 10.79 |
| 168 | 34.74 | 127.74 | 34.63 | 127.65 | 14.22 |
| 169 | 34.69 | 125.45 | 34.65 | 125.55 | 10.04 |
| 170 | 34.40 | 126.70 | 34.40 | 126.75 | 4.26 |
| 172 | 35.35 | 126.60 | 35.39 | 126.46 | 13.75 |
| 174 | 35.02 | 127.37 | 35.13 | 127.36 | 12.58 |
| 175 | 34.47 | 126.32 | 34.40 | 126.45 | 13.98 |
| 184 | 33.51 | 126.53 | 33.41 | 126.44 | 13.76 |
| 185 | 33.29 | 126.16 | 33.42 | 126.15 | 13.56 |
| 188 | 33.39 | 126.88 | 33.41 | 126.74 | 13.59 |
| 189 | 33.25 | 126.57 | 33.17 | 126.44 | 14.48 |
| 192 | 35.16 | 128.04 | 35.12 | 127.97 | 8.05 |
| 201 | 37.71 | 126.45 | 37.63 | 126.47 | 9.43 |
| 202 | 37.49 | 127.49 | 37.37 | 127.41 | 15.43 |
| 203 | 37.26 | 127.48 | 37.37 | 127.41 | 13.24 |
| 211 | 38.06 | 128.17 | 38.10 | 128.06 | 10.55 |
| 212 | 37.68 | 127.88 | 37.61 | 127.73 | 15.51 |
| 216 | 37.17 | 128.99 | 37.08 | 128.96 | 9.81 |
| 217 | 37.38 | 128.65 | 37.34 | 128.66 | 4.60 |
| 221 | 37.16 | 128.19 | 37.10 | 128.34 | 14.45 |
| 226 | 36.49 | 127.73 | 36.37 | 127.70 | 13.72 |
| 232 | 36.76 | 127.29 | 36.87 | 127.40 | 15.29 |
| 235 | 36.33 | 126.56 | 36.38 | 126.46 | 10.40 |
| 236 | 36.27 | 126.92 | 36.38 | 126.77 | 17.93 |
| 238 | 36.11 | 127.48 | 36.12 | 127.38 | 9.15 |
| 243 | 35.73 | 126.72 | 35.64 | 126.76 | 11.27 |
| 244 | 35.61 | 127.29 | 35.63 | 127.37 | 8.03 |
| 245 | 35.56 | 126.87 | 35.64 | 126.76 | 12.33 |
| 247 | 35.40 | 127.40 | 35.38 | 127.37 | 3.60 |
| 248 | 35.66 | 127.52 | 35.63 | 127.37 | 13.75 |
| 251 | 35.43 | 126.70 | 35.39 | 126.76 | 7.14 |
| 252 | 35.28 | 126.48 | 35.39 | 126.46 | 11.92 |
| 253 | 35.23 | 128.89 | 35.35 | 128.88 | 13.14 |
| 254 | 35.37 | 127.13 | 35.38 | 127.06 | 6.11 |
| 255 | 35.23 | 128.67 | 35.11 | 128.57 | 15.88 |
| 257 | 35.31 | 129.02 | 35.35 | 128.88 | 13.08 |
| 258 | 34.76 | 127.21 | 34.64 | 127.35 | 18.80 |
| 259 | 34.63 | 126.77 | 34.65 | 126.75 | 2.70 |
| 260 | 34.69 | 126.92 | 34.64 | 127.05 | 13.10 |
| 261 | 34.55 | 126.57 | 34.65 | 126.45 | 15.03 |
| 262 | 34.62 | 127.28 | 34.64 | 127.35 | 7.31 |
| 263 | 35.32 | 128.29 | 35.36 | 128.28 | 4.70 |
| 264 | 35.51 | 127.75 | 35.62 | 127.68 | 13.96 |
| 266 | 34.94 | 127.69 | 34.88 | 127.66 | 7.48 |
| 268 | 34.47 | 126.26 | 34.40 | 126.15 | 12.79 |
| 271 | 36.94 | 128.91 | 36.84 | 128.95 | 12.48 |
| 272 | 36.87 | 128.52 | 36.84 | 128.64 | 11.48 |
| 273 | 36.63 | 128.15 | 36.61 | 128.01 | 12.24 |
| 276 | 36.43 | 129.04 | 36.34 | 128.93 | 14.45 |
| 277 | 36.53 | 129.41 | 36.57 | 129.56 | 13.88 |
| 278 | 36.36 | 128.69 | 36.35 | 128.62 | 6.06 |
| 279 | 36.13 | 128.32 | 36.11 | 128.30 | 2.89 |
| 281 | 35.98 | 128.95 | 36.09 | 128.92 | 13.01 |
| 283 | 35.82 | 129.20 | 35.83 | 129.21 | 2.10 |
| 284 | 35.67 | 127.91 | 35.62 | 127.98 | 8.55 |
| 285 | 35.57 | 128.17 | 35.61 | 128.29 | 11.75 |
| 288 | 35.49 | 128.74 | 35.60 | 128.90 | 17.95 |
| 289 | 35.41 | 127.88 | 35.37 | 127.97 | 9.83 |
| 294 | 34.89 | 128.60 | 34.86 | 128.56 | 4.86 |
| 295 | 34.82 | 127.93 | 34.88 | 127.96 | 7.24 |
| 90 | 38.25 | 128.56 | 38.34 | 128.70 | 15.59 |
| 95 | 38.15 | 127.30 | 38.11 | 127.43 | 11.31 |
| 98 | 37.90 | 127.06 | 37.87 | 127.11 | 5.32 |
| 99 | 37.89 | 126.77 | 37.87 | 126.79 | 2.52 |

# Figures

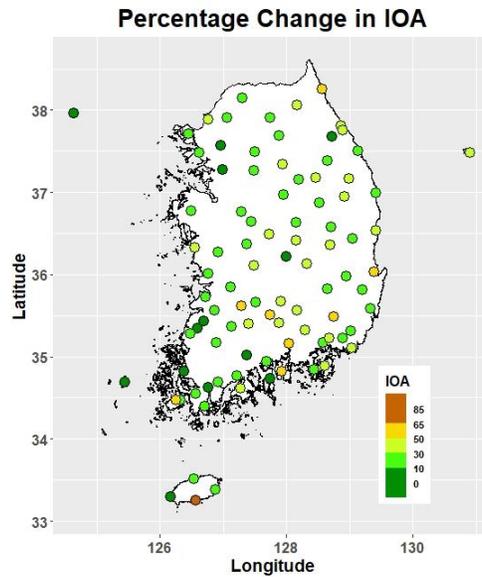

**Figure S1.** Percentage change in the IOA from the WRF to the Weather-AI models for wind speed forecasting

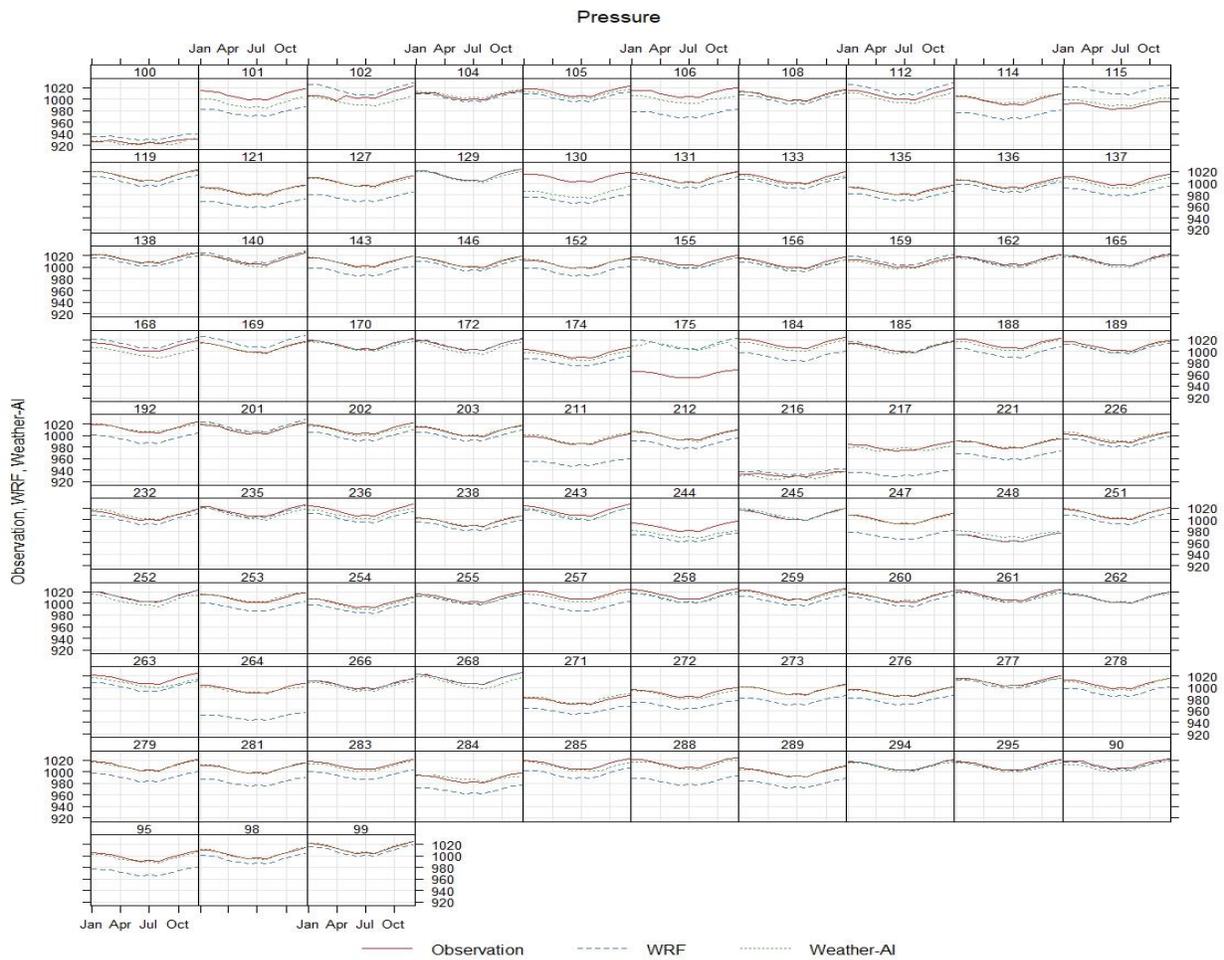

**Figure S2.** Monthly mean of surface pressure (in hPa) of stations in South Korea. The X-axis represents the months for the year 2018 and Y-axis represents surface pressure.